\documentstyle[12pt]{article}
\begin{document}
\title{\bf The imprints of nonextensive statistical mechanics
in high energy collisions\footnote{Invited talk presented by G.Wilk
at {\it International Workshop on Classical and Quantum Complexity
and Non-extensive Thermodynamics} (Complexity 2000), Denton,
University of North Texas, April 3-6, 2000.} }
\author{G.Wilk$^{1}$\thanks{e-mail: wilk@fuw.edu.pl} and Z.W\l
odarczyk$^{2}$ \thanks{e-mail: wlod@pu.kielce.pl}\\[2ex] 
$^1${\it The Andrzej So\l tan Institute for Nuclear Studies}\\
    {\it Ho\.za 69; 00-689 Warsaw, Poland}\\
$^2${\it Institute of Physics, Pedagogical University}\\
    {\it  Konopnickiej 15; 25-405 Kielce, Poland}}  
\date{\today}
\maketitle

\begin{abstract}
To describe high energy collisions one widely uses thermodynamical 
methods and concepts which follow the classical Boltzmann-Gibbs (BG) 
approach. In many cases, however, either some deviations from the 
expected behaviour are observed experimentally or it is known that 
the conditions necessary for BG to apply are satisfied only 
approximately. In other branches of physics where such situations 
are ubiquitous, the popular remedy is to resort, instead, to the so 
called nonextensive statistics, the most popular example of which
is Tsallis statistics. We shall provide here an overview of possible
imprints of non-extensitivity existing both in high energy cosmic ray
physics and in multiparticle production processes in hadronic
collisions, in particular in heavy ion collisions. Some novel
proposition for the interpretation of the nonextensitivity parameter
$q$ present in such circumstances will be discussed in more detail.\\

[3ex]

\end{abstract}

\newpage
\section{Introduction}

It has been realized for some time already that in many places of
science there are phenomena which clearly indicate the existence of
some degree of nonextensitivity (understood in the thermodynamical
sense). They include all situations characterized by long-range
interactions, long-range microscopic memory and space-time (and
phase space as well) (multi)fractal structure of the process. Such
anomalous (from the point of view of the standard thermodynamics and
Boltzmann-Gibbs statistics) systems are found to be best described in
terms of generalised, nonextensive thermostatistics, the most 
popular and explored example of which is the so called Tsallis statistics
which is characterised by the nonextensitivity parameter $q$ \cite{T}.\\

To make our presentation self-contained we shall first provide the basic
formulas (refering to cf. \cite{T} and references therein for a
thorough discussion of all possible aspects of nonextensivity).
Everything is based on a generalized entropic form depending on a single
parameter (entropic index) $q$ in such a way that for $q \rightarrow
1$ it gives the normal BG entropy: 
\begin{equation}
S_q = -\, \frac{1}{1 - q} \left( 1\, -\, \sum_i p_i^q \right)
        \qquad \stackrel{q \rightarrow 1}{\Longrightarrow} \qquad 
        S_{BG}\, =\, -\, \sum_i p_i\, \ln p_i .
 \label{eq:Def}
\end{equation}
The $S_q$ is nonextensive in the sense that
\begin{equation}
S_q(A+B)\, = \, S_q(A)\, +\, S_q(B)\, + (1-q)\, S_q(A)S_q(B) ,
         \label{eq:Nonex}
\end{equation}         
where $A$ and $B$ are two independent systems in the usual sense,
i.e., $p_{ij}(A+B)=p_i(A)p_j(B)$. In this sense the entropic index
$q$ is also a measure of the nonextensivity in the system. Using the
usual procedure of information theory when looking for the most
probable and least biased (normalized) probability distribution of
some events $x$ subjected to a single constraint in the form of
unnormalized $q$-expectation value $<A>_q = \int dx A(x)p(x)^q$
\cite{FOOT0}, one gets immediately, from the maximization of the
entropy $S_q$ the known expression 
\begin{equation}
p_q(x)\, =\, \frac{1}{Z_q}\left[ 1\, -\, (1 - q)
\alpha A(x)\right]^{\frac{1}{1-q}} \qquad \stackrel{q\rightarrow
1}{\Longrightarrow}\qquad \frac{1}{Z} \exp\left( - \alpha A(x)\right)
. \label{eq:Prob} 
\end{equation} 
The Lagrange multiplier $\alpha$ can be expressed in terms of the
$<A>_q$ from the imposed constraint and $Z_q$ from the normalization
condition. Notice that whereas in the extensive, i.e., $q=1$ case, all
values of $x\in (0,\infty)$ are admissible, for nonextensive case of
$q\neq 1$ we have restrictions so that $[1 - (1-q)\alpha A(x)]$ is
positive.\\ 

Our presentation will be devoted to a very limited subject of high
energy collisions. We shall attempt to overview the probable imprints
of nonextensivity in high energy collisions where under this term we
understand both multiparticle production processes taking place in
cosmic ray experiments and those observed in accelerator experiments.
\\ 

In cosmic ray experiments one encounters routinely a cascade
processes, both in the atmosphere and in emulsion chambers serving as
detectors. The former originate during the passage of the primordial
cosmic rays through the atmosphere with subsequent collisions and
dissipation of energy (they are known as the so called Extensive Air
Showers) - this is typical stochastic process not very much sensitive
to details of the elementary interactions \cite{CASC}. The latter are
connected with the actual detection process taking place in special
emulsion chambers exposed in many places of the Earth, usually at
high altitudes. Some of their characteristics do depend on
details of the interaction process. Because in both
cases one encounters formulae of the type (\ref{eq:Prob}), they are
{\it a priori} sensitive to the possible nonextensivity of such
processes. \\ 

The accelerator high energy collisions are usually connected with
production of large number of secondaries (mostly $\pi$ and $K$
mesons). The strong interactions involved here make their detail
descriprion {\it from first principles} impossible and one is forced
to turn to phenomenological models of various kinds. The most
economical (as far as the number of parameters is concerned) are
thermodynamical and statistical models which have been, in fact, in use
since the beginning of this field of research. It should be stressed
that the very first successful phenomenological model of
multiparticle production, the so called Landau Hydrodynamical Model,
was proposed already in the pre-accelerator era of 1953 and used in
the analysis of specific multiparticle data taken from cosmic ray
interactions \cite{LANDAU}. This model paved a way to more
sophisticated statistical models now in use, mainly in high energy
heavy ion collisions which are believed to lead to the production of
new state of matter, the Quark Gluon Plasma (QGP) \cite{QGP}. \\ 

To recapitulate: in both cosmic ray and accelerator experiments one
observes multiparticle production. However, in the cosmic ray case one
also deals with propagation of the secondaries originating in
multiple production processes at a given point and with their
subsequent secondary interactions (also of multiparticle type), i.e.,
with a full fledged cascade process \cite{CASC}.\\

In Section 2 the possible traces of nonextensivity apparently seen in
some cosmic ray experiments \cite{CR} with emulsion chambers will be
presented along with similar effects seen in elementary $e^+e^-$ \cite{BCM}
and hadronic collisions and in the collisions of nuclei in accelerators 
\cite{ALQ,UWW}. In all of them distributions given by Eq. (\ref{eq:Prob}) 
are observed. Section 3 will be devoted to disscusion on possible
hints emerging from the occupation number distributions $<n>_q$ 
\cite{ALQ,UWW}. In Section 4 we shall propose a novel interpretation
of the entropic index $q$ \cite{WW} encountered in Sec. 2. The last
Section contains final remarks together with a list of other possible
hints of nonextensivity not discussed in detail here (including 
topics from QGP physics \cite{JR} and nonextensivity
manifested in the statistics of the quantum states produced in the
scattering process such as $\pi-$nucleon or $\pi-$nucleus scatterings
\cite{II}).\\

\section{Traces of nonextensivity in nonexponential distributions}

Our encounter with the notion of nonextensivity started when we 
realised that one of our previous results concerning the occurence
of the so called {\it long flying component} in the propagation of
the initial flux of incoming cosmic ray particles (mostly nucleons)
\cite{WWCR} can be interpreted as yet another manifestation of the
L\'evy distribution (\ref{eq:Prob}) \cite{CR}. To be more specific
let us briefly summarize the result of \cite{WWCR}. We have analysed,
there, the distribution of cascade starting points in the extrathick lead
chamber of the Pamir experiment. The corresponding data points for the
number of cascades originating at depth $T$ (measured in cascade
units, $1$ c.u. $=6.4{\rm g}/{\rm cm}^2=0.56{\rm cm}$) are shown in
Fig. 1. Whereas at small depths (up to $\sim 60$ cm of lead) we
observe the usual absorption of hadrons as given by the simple
exponential formula
\begin{equation}
\frac{dN}{dT}\, =\, {\rm const}\cdot \exp\left(-
      \frac{T}{\lambda}\right) , \label{eq:Q1}
\end{equation}
at biggest thickness there is, against all expectations, noticeably
excess of experimental points above the simple extrapolation of 
small-depth data. The observed discrepancy means that original hadrons tend to
fly longer without interaction (that is why a term {\it long flying
component} is coined for this type of phenomenon).\\   

In \cite{WWCR} we have argued that the observed effect can be just
another manifestation of the fluctuation of the corresponding
hadronic cross section $\sigma = Am_N\frac{1}{\lambda}$ (where $A$
denotes the mass number of the target and $m_N$ is the mass of the
nucleon, such a possibility is widely discussed in the literature and
observed in diffraction dissociation experiments on accelerators, cf.
\cite{WWCR} for details and references). It turned out that
fluctuations of this cross section (i.e., in effect, fluctuations of
the quantity $1/\lambda$) with relative variance
\begin{equation}
\omega\, =\, \frac{\langle \sigma^2\rangle - \langle \sigma \rangle ^2}
          {\langle \sigma \rangle ^2}\, \geq 0.2 \label{eq:omega}
\end{equation}
allow to describe the observed effect. \\

It turns out \cite{CR} that the same data can be fitted by the
nonextensive formula 
\begin{equation}
\frac{dN}{dT}\, =\, {\rm const}\cdot \left[1\, -\,
      (1-q)\frac{T}{\lambda}\right]^{\frac{1}{1-q}} \label{eq:Q13} 
\end{equation}
with parameter $q=1.3$ (in both cases $\lambda = 18.85\pm 0.66$ in
$c.u.$ defined above), cf. Fig. 1 \cite{FOOT1}.\\

Similar example is also known in heavy ion collisions \cite{UWW}
(see also \cite{ALQ}). It turns out that distributions of transverse
momenta of secondaries produced in nuclear collisions at high
energies (transverse with respect to the collision axis given by the
direction of the colliding objects in the center of mass frame)
$dN(p_T)/dp_T$ are best described by a slightly nonexponential
function of the type 
\begin{equation}
\frac{dN(p_T)}{dp_T}\, =\, {\rm const}\cdot \left[ 1\, - (1 - q)
\frac{\sqrt{m^2 + p^2_T}}{kT}\right]^{\frac{1}{1-q}}\qquad
\stackrel{q\rightarrow 1}{\Longrightarrow}\qquad {\rm const}\cdot
\exp\left( - \frac{\sqrt{m^2 + p^2_T}}{kT}\right) . \label{eq:Pt}
\end{equation}
Here $m$ is the mass of produced particle, $k$ is the Boltzmann
constant (which we shall, in what follows, put equal unity) and $T$ is,
for the $q=1$ case, the {\it temperature} of the reaction considered
(or, rather, the temperature of the hadronic system produced). 
In fact, precisely from such exponential fits to transverse masses
$m_T = \sqrt{m^2 + p_T^2}$ one infers information about the temperature
$T$. Therefore, any deviation from the exponential behaviour of such
distributions are always under detailed scrutiny in which one is
searching for the possible causes. In \cite{ALQ} it was suggested
that the extreme conditions of high density and temperature occuring
in ultrarelativistic heavy ion collisions can lead to memory effects
and long-range colour interactions and to the presence of
non-Markovian processes in the corresponding kinetic equations
\cite{NONM}. \\

As it is seen in Fig. 2 one indeed finds \cite{ALQ,UWW} a small deviation
from the exponential behaviour (on the level $q=1.015$). As we shall
demonstrate in Section 4 it can, however, lead to quite dramatic
effects. It was also shown in \cite{ALQ} that to first order in
$|q-1|$ the generalized slope becomes the quantity
\begin{equation}
T_q\, =\, T\, +\, (q-1) m_T . \label{eq:TQ}
\end{equation}
with $T$ being temperature of {\it a purely thermal source}.
This should be contrasted with the empirical relation for the slope
parameter $T$, from which the freeze-out temperature (at which
hadrons are created from the QGP) $T_f$ is then deduced,
\begin{equation}
T\, =\, T_f\, +\, m\langle v_{\perp}\rangle^2 . \label{eq:Tf}
\end{equation}
The $\langle v_{\perp}\rangle$ is a fit parameter usually identified with
the average collective (transverse) flow velocity of the hadrons
being produced. In (\ref{eq:TQ}) one has, instead, {\it a purely
thermal source} experiencing a kind of blue shift at high $m_T$
(actually increasing with $m_T$). The nonextensivity parameter $q$
accounts here for all possibilities one can find in \cite{NONM} and
could, therefore, be regarded as a new way of presenting experimental
results with $q\neq 1$ signaling that there is something going on in
the collision that prevents it to from being exactly thermal-like in the
ordinary sense mentioned above.\\

We shall proceed, now, to two other examples of possible
nonexponential distributions. First is the attempt \cite{BCM} to fit
the energy spectra in both the longitudinal and transverse momenta of
particles produced in the $e^+e^-$ annihilation processes at high
energies. Those are, contrary to the previous example, the most
elementary high energy multiparticle production processes.
The initial $e^+e^-$ pair annihilates to a virtual photon which
subsequently gives rise to a (highly excited) quark-antiquark pair.
They in turn develop a complex hadronization process related to the
long-distance (strong coupling) regime of Quantum Chromodynamic (QCD).
Usually being described in terms of the so called string model
\cite{LUND} it admits also, for low energies, a kind of
thermodynamical equilibrium approach \cite{BCM}. However, it turns
out that when going to higher energies one cannot keep the
temperature $T_0$ inferred from the $p_T$ distributions constant,
invalidating therefore the whole concept. On the other hand, using
instead the nonextensive power-like ($q$-dependent) distribution, one
can write the following transverse momentum distribution \cite{BCM}:
\begin{equation}
\frac{1}{\sigma}\frac{d\sigma}{dp_T}\, =\, {\rm const}\cdot p_T\, 
 \int^{\infty}_0\, dp_L\, \left[ 1\, -\, (1 - q)\,
 \frac{\sqrt{p^2_L + m^2 + p_T^2}}{T_0}\right]^{\frac{1}{1-q}} \label{eq:HAGq}
\end{equation}
(where $p_L$ is longitudinal momentum of secondary particle of mass $m$
and $q$ is the entropic index). Keeping the temperature $T_0$ essentially
constant and changing only $q$ one can now fit data extremely well, cf.
Fig. 3 \cite{BCM}. In this example the $q\neq 1$ is then regarded 
as a manifestation of
nonextensivity arising in the hadronization process in which quarks
and gluons combine together forming hadrons, a process which involves
long range correlations in the phase space. Actually, this
observation has general validity and applies to all production
processes discussed here as well. It applies, most probably, also to 
a pure elastic quantum scattering processes discussed in \cite{II}.\\

The final example from this category deals with the most probable
rapidity distributions for produced secondaries. Suppose that we have
an excited object of mass $M$ which hadronizes into $N$ secondaries
of transverse mass $m_T$ each. For simplicity we shall consider only
one-dimensional hadronization in which transverse momenta are hidden
in the mean $\langle p_T \rangle$ parameter, which is kept
constant and enters $m_T$. In this case the observable distribution of
interest is $f(y)=\frac{1}{N}\frac{dN}{dy}$, i.e., a
distribution in rapidity defined as $y=\frac{1}{2}\ln [(E + p_L)/(E -
p_L)]$. Some time ago it was shown, using the maximization of the
information entropy in the Shannon form (i.e., in fact the BG one), that
knowing only the energy $M$ and multiplicity $N$ one has to expect
that \cite{MAXENT} 
\begin{equation}
f(y)\, =\, \frac{1}{Z(M,N)} \exp\left[ - \beta(M,N)\cdot m_T \cosh y
\right] , \label{eq:MAXENT}
\end{equation}
where $Z(M,N)$ comes from the normalization of $f(y)$ and the
Lagrange multiplier $\beta(M,N)$ from the energy conservation
constraint. Of special interest to us is the fact that for some
values of the mean energy $M/N$ (i.e., for some values of $N$ for a given
mass $M$) $\beta$ can be zero and even negative. Actually it can be
shown that $\beta \geq 0$ only if
\begin{equation}
N\, \geq \, N_0\, \simeq 2 \ln\frac{M}{m_T} . \label{eq:Feynman}
\end{equation}
This statement invalidates the widely assumed, on different occassions,
the so-called {\it Feynman scaling} hypothesis, namely, that in high
energy collisions one should expect $f(y) \simeq {\rm const}$ (cf.
\cite{MAXENT} for more information). Such hypothesis is clearly
incompatible with the experimentally observed fact that the mean
multiplicity $<N>$ grows with energy faster than $\ln M$, more like
$<N> \sim M^{0.4-0.5}$). However, using Tsallis $q$-entropy, instead
of BG one obtains \cite{UWWF}
\begin{equation}
f_q(y)\, =\, \frac{1}{Z_q(M,N)} \left[ 1\, -\, (1 - q) \beta_q(M,N)\cdot m_T
                \cosh y \right]^{\frac{1}{1-q}} \label{eq:MAXENTq}
\end{equation}
with $\beta_q(M,N) \geq 0$ for  
\begin{equation}
N\, \geq N_{q0}\, \simeq 2 \left(\ln \frac{M}{m_T}\right)^q . 
           \label{eq:Feynmanq}
\end{equation}
It is, then, obvious that for $q>1$ one can indeed accommodate, at a given
rapidity interval, more particles with $f(y) = f_q(y)$ being constant.
In a sense one can think of a kind of {\it Feynman $q$-scaling}
here (which would generalize the usual one and describe, in terms of
the parameter $q$, what is usually called a violation of the Feynman
scaling hypothesis \cite{UWWFa}). In Fig. 4 this effect is clearly
demonstrated for $M=100$ GeV (and $m_T=0.4$ GeV). While in Fig. 4a
$N=10$ is chosen in such a way as to have $\beta(100,10) \simeq
0$ (notice that in this case $\beta_{q=0.7} > 0$ whereas
$\beta_{q=1.3} < 0)$, in Fig. 4b one has, instead,
$\beta_{q=1.3}(100,20) =0$ (and remaining $\beta_{q=1} > 0$ and
$\beta_{q=0.7} > 0$). Fig 4c shows, for comparison, the case with $N >
N_0$ for which all $\beta_q > 1$.\\

\section{Traces of nonextensivity in the mean occupation numbers
$n_q$ } 

Another place where nonextensivity enters in a natural way is the
mean occupation numbers generalizing the Bose-Einstein or Fermi-Dirac
ones to a $q\neq 1$ case. Whereas the single particle distribution
function is obtained in the usual procedure of maximizing the Tsallis
entropy under the constraints of given average internal energy and
number of particles, the mean occupation numbers $\langle n \rangle_q$ are not
available in analytical formula for any $q$. Only in the dilute gas
approximation and for small deviations of $q$ from unity can one
express them in a simple analytical form \cite{QSTAT}
\begin{equation}
\langle n\rangle_q\, =\, \left\{ \left[1\, +\,
(q-1)\beta(E-\mu)\right]^{1/(q-1)}\pm1\right\}^{-1} , \label{eq:QSTAT}
\end{equation}
where $\beta=1/kT$, $\mu$ is the chemical potential and the $+/-$ sign
applies to fermions/bosons. Notice that in the limit $q \rightarrow
1$ (extensive statistics) one recovers the conventional Fermi-Dirac
and Bose-Einstein distributions. What will interest us here are the
generalized particle fluctuations,
\begin{equation}
\langle \Delta n^2\rangle_q \, \equiv \, \frac{1}{\beta} \frac{\partial
        \langle n\rangle_q}{\partial \mu}\, =\, 
        \frac{\langle n\rangle_q}{1 + (q-1)\beta(E - \mu)}
        \left( 1 \mp \langle n\rangle_q\right), \label{eq:Fluq}
\end{equation}
where $E=\sqrt{m^2 + p^2}$. That is because, so far, these formulas
have been applied to study the fluctuation pattern expected in heavy ion 
collisions \cite{ALQ,UWW} in measurements performed on an
event-by-event basis. Notice that the denominator occuring
in (\ref{eq:Fluq}) modulates in a novel and specific way the usual
pattern of fluctuations for the $q=1$ case \cite{ALQ}.\\        

Event-by-event fluctuations can be used as a valuable source of
information on the dynamics of heavy-ion collisions. However, there
is the problem of how to disentangle the {\it dynamical} fluctuations of
interest from the {\it trivial} geometrical ones due to the impact
parameter variation (resulting in the different number of nucleons
participating in a given event). To solve this problem it was
proposed in \cite{GM} to use the following measure of fluctuations or
correlations: 
\begin{equation}
\Phi_x\, =\, \sqrt{\frac{\left\langle Z^2\right\rangle}{\langle N\rangle}}
          \, -\, \sqrt{\bar{z^2}} 
          \qquad {\rm where}\qquad
          Z\, =\, \sum^N_{i=1}\, z_i .
           \label{eq:FI}
\end{equation}
Here $z_i = x_i - \bar{x}$ where $\bar{x}$ denotes the mean value of
the observable $x$ calculated for all particles from all events (the
so called inclusive mean) and $N$ is the number of particles analysed
in the event. In (\ref{eq:FI}) $\langle N\rangle$ and $\langle
Z^2\rangle$ are averages of event-by-event observables over all 
events whereas the last term is the square root of the second
moment of the inclusive $z$ distribution. $\Phi$ equals zero when the
correlations are entirely absent. On the other hand, it is
constructed in such a way as to be exactly the same for
nucleon-nucleon and nucleus-nucleus collisions if the latter is a
simple superposition of the former. The $\Phi$-measure has been
successfully applied to the experimental data (cf. \cite{MN})
and the fluctuations of transverse momentum $p_T$, which are observed
in nucleon-nucleon collisions, have been found to be significantly
reduced in the central Pb-Pb collisions at $158$ GeV per nucleon. 
Its nonextensive extension has been analysed in \cite{ALQ}
where it was found that the usual correlations are increased for
$q<1$ and decreased for $q>1$. This time the measure $\Phi$ can
became positive or negative for both fermions and bosons depending on
the value of the entropic index $q$. Actually in \cite{ALQ} it was
found that for $T=140$ MeV and $\mu =0$ the $\Phi$-measure vanishes
for $q=1.015$.\\ 

The clear prediction of \cite{ALQ}, which can consist a subject for
experimental verification, is that the $p_T$-dependence of partial
contributions to $\Phi_{p_T}$ should become negative for
$p_T > 0.5$ GeV. Notice that in this regime the $\Phi$-measure should
already be free from contaminations from resonance decays and,
therefore, the experimental confirmation of this prediction would
provide a strong signal for the nonextensivity present in relativistic
heavy ion collisions.\\

The $\Phi$-measure is applicable not only to fluctuations of kinematical
quantities such as $p_T$ but also to the azimuthal \cite{MNPHI} and  
chemical fluctuations as well. The latter were analysed in \cite{MN1}
for the normal statistics and in \cite{UWW} for the nonextensive one.
The representive sample of results is shown in Fig. 5 for
$q=1.015$ mentioned above. For simplicity we have restricted
ourselves, here, only to comparison with results of \cite{MN1} without
resonances \cite{FOOT6}. Actually, one expects that for a given $q$
fluctuations should grow with the mass of detected particle - this
observation provides yet another possibility for experimental
verification of the nonextensivity concept.\\

\section{Nonextensivity parameter $q$ as a measure of fluctuations}

The general picture emerging from the previous discussion is that one
can account very economically (by introducing only one new parameter
$q$) and adequately (by using nonextensive formulas emerging from
Tsallis statistics with entropic index $q$) for a number of
observations deviating from the normal BG approach. The question of
the possible meaning of entropic index in these cases is therefore
very natural. For the cases discussed in Section 2 we would like to
propose that $q$ is connected with fluctuations present in the system
under investigation. Notice that common feature of the 
first two examples in Section 2 is that they both are described by
the powerlike distribution of the type
\begin{equation}
L_q(\varepsilon)\, =\, C_q\, \left[\, 1\, -  
                   \, (1\, -\, q)\, \frac{\varepsilon}{\lambda}\,
                   \right]^{\frac{1}{1 - q}}. \label{eq:T}
\end{equation}                   
As mentioned before, the cosmic ray example was originally explained
\cite{WWCR} by the apparent fluctuation of the mean free path
parameter $\lambda$ in the corresponding exponential formula
(\ref{eq:Q1}). It is then natural to expect that these fluctuations
(which were so far described in \cite{WWCR} only numerically by means
of Monte Carlo simulations) should be formulated in such a way as to
result in eq. (\ref{eq:T}) with a parameter $q$. The same should be
also true for the heavy ion collision example. Actually, this example is
even more important and interesting because of the long and still vivid
discussion on the possible dynamics of temperature fluctuations
\cite{L,LS,FLUCT1,FLUCT2} and because of its connection with the problem
of QGP production in heavy ion collisions \cite{LS,FLUCT2}. We
shall, therefore, treat both cases as representing the same class of
fluctuation phenomena and claim that {\it the parameter $q$ is a measure
of fluctuations} present in the L\'evy-type distributions
(\ref{eq:T}) describing the particular process under consideration.\\  
  
To demonstrate this conjecture let us analyse the influence
of fluctuations of the parameter $1/\lambda$ present in the exponential
formula $L_{q=1}(\varepsilon) \sim \exp(-\varepsilon/\lambda)$. 
Our aim will be to deduce the form of a function
$f(1/\lambda)$ which transforms the exponential distribution to a
power-like L\'evy distribution (\ref{eq:T}) and which describes
fluctuations about the mean value $1/\lambda_0$. Although in both
examples considered above the data preferred $q>1$, we shall discuss
$q<1$ case as well. In the $q>1$ case, where $\varepsilon \in (0,
\infty)$, one has, 
\begin{equation}
L_{q>1}(\varepsilon;\lambda_0)\, =\, C_q\, 
\left( 1\, +\, \frac{\varepsilon}{\lambda_0}\, \frac{1}{\alpha}\right)^{-a}\, =\,
C_q\, \int^{\infty}_0\, \exp\left( - \frac{\varepsilon}{\lambda}\right)\,
f\left(\frac{1}{\lambda}\right)\, d\left(\frac{1}{\lambda}\right)
\label{eq:DEF} 
\end{equation}
where $\alpha = \frac{1}{q-1}$. Writing the following representation
of the Euler gamma function \cite{FOOT2},
\begin{equation}
\left( 1\, +\, \frac{\varepsilon}{\lambda_0}\, \frac{1}{\alpha}\right)^{-a}\, =\,
\frac{1}{\Gamma(\alpha)}\, \int^{\infty}_0\, d\eta\, 
      {\eta}^{\alpha - 1}\, \exp\left[ - \eta\, 
      \left(1\, +\, \frac{\varepsilon}{\lambda_0}\,
       \frac{1}{\alpha}\right)\right] ,
 \label{eq:GF}
\end{equation}
and changing variables under the integral to $\eta=\alpha
\frac{\lambda_0}{\lambda}$, one obtains eq. (\ref{eq:DEF}) with 
$f(1/\lambda)$ given by the following gamma distribution:
\begin{equation}
f_{q>1}\left(\frac{1}{\lambda}\right)\, =\,
f_{\alpha}\left(\frac{1}{\lambda},\frac{1}{\lambda_0}\right)\, =\,
\frac{\mu}{\Gamma(\alpha)}\,
\left(\frac{\mu}{\lambda}\right)^{\alpha-1}\, \exp\left(
- \frac{\mu}{\lambda}\right) \label{eq:F}
\end{equation}
with $\mu = \alpha \lambda_0$ and with mean value and variation in the form:
\begin{equation}
\left\langle \frac{1}{\lambda}\right\rangle \, =\,
 \frac{1}{\lambda_0} \qquad {\rm and}\qquad
\left\langle \left(\frac{1}{\lambda}\right)^2\right\rangle\, -\, 
\left\langle\frac{1}{\lambda}\right\rangle^2\, =\, 
\frac{1}{\alpha\, \lambda_0^2} . \label{eq:MEANVAR}
\end{equation}
Notice that, with increasing $\alpha$ the variance (\ref{eq:MEANVAR})
decreases and asymptotically (for $\alpha \rightarrow \infty$, i.e,
for $q\rightarrow 1$) the gamma distribution (\ref{eq:F}) becomes
a delta function, $f_{q>1}(1/\lambda)=\delta (\lambda - \lambda_0)$. The
relative variance for this distribution is given by 
\begin{equation}
\omega\, =\, \frac{\left\langle\left(\frac{1}{\lambda}\right)^2\right\rangle\, 
 -\, \left\langle\frac{1}{\lambda}\right\rangle^2}
 {\left\langle \frac{1}{\lambda}\right\rangle^2}\, =\,
\frac{1}{\alpha}\, =\, q\, -\, 1 . \label{eq:PROOF}
\end{equation}

For the $q<1$ case $\varepsilon$ is limited to $\varepsilon \in
[0,\lambda_0/(1-q)]$. Proceeding in the same way as before, i.e.,
making use of the following representation of the Euler gamma function
(where $\alpha' = - \alpha = \frac{1}{1-q}$)
\begin{equation}
\left[1\, -\, \frac{\varepsilon}{\alpha' \lambda_0}\right]^{\alpha'}\, =\,
\left(\frac{\alpha' \lambda_0}{\alpha' \lambda_0 -
\varepsilon}\right)^{-\alpha'}\, 
=\, \frac{1}{\Gamma(\alpha')}\, \int^{\infty}_0\, d\eta\, 
    \eta^{\alpha' - 1}\, \exp\left[ - \eta\, \left(1\, +\, 
    \frac{\varepsilon}{\alpha' \lambda_0 - \varepsilon}\right)\right]
      , \label{eq:EGF}
\end{equation}
and changing variables under the integral to $\eta = \frac{\alpha'
\lambda_0 - \varepsilon}{\lambda}$, we obtain
$L_{q<1}(\varepsilon;\lambda_0)$ in the form of eq. (\ref{eq:DEF}) but
with $\alpha \rightarrow -\alpha'$ and with the respective $f(1/\lambda)
= f_{q<1}(1/\lambda)$ given now by the same gamma distribution as in
(\ref{eq:F}) but this time with $\alpha \rightarrow \alpha'$ and
$\mu = \mu(\varepsilon) = \alpha' \lambda_0 - \varepsilon$. Contrary to the
$q>1$ case, this time the fluctuations depend on the value of the
variable in question, i.e., the mean value and variance are now both
$\varepsilon$-dependent:  
\begin{equation}
\left\langle \frac{1}{\lambda}\right\rangle\, =\, \frac{1}{\lambda_0 -
\frac{\varepsilon}{\alpha'}}\qquad {\rm and}\qquad \left\langle
\left(\frac{1}{\lambda}\right)^2\right\rangle\, -\,
\left\langle\frac{1}{\lambda}\right\rangle^2\, =\, \frac{1}{\alpha'}\cdot
\frac{1}{\left(\lambda_0 - \frac{\varepsilon}{\alpha'}\right)^2} .
\label{eq:MV}
\end{equation}
However, the relative variance
\begin{equation}
\omega\, =\,  \frac{\left\langle\left(\frac{1}{\lambda}\right)^2\right\rangle\,
       -\, \left\langle\frac{1}{\lambda}\right\rangle^2}
       {\left\langle \frac{1}{\lambda}\right\rangle^2}\, 
       =\, \frac{1}{\alpha'}\, =\, 1\, -\, q , \label{eq:RESULT}
\end{equation}
remains $\varepsilon$-independent and depends only on the parameter
$q$. As above the resulting gamma distribution becomes a delta
function, $f_{q<1}(1/\lambda)=\delta (\lambda - \lambda_0)$, for
$\alpha' \rightarrow \infty$, i.e., for $q\rightarrow 1$.\\  

This completes the proof of our conjecture. The nonextensivity parameter $q$
in the $L_q(\varepsilon)$ distributions can, indeed, be expressed by the
relative variance $\omega$ of fluctuations of the parameter $1/\lambda$
in the distribution $L_{q=1}(\varepsilon)$:
\begin{equation}
q = 1 \, \pm \, \omega       \label{eq:QG1}
\end{equation}
for the $q>1$ ($+$) and $q<1$ ($-$) cases.\\

Concerning transverse momentum distributions in
heavy ion collisions, $dN(p_T)/dp_T$, it is interesting to notice
that the relatively small value $q \simeq 1.015$ of the nonextensive
parameter obtained there \cite{ALQ,UWW}, if interpreted in the same
spirit as above, indicates that rather large relative fluctuations of
temperature, of the order of $\Delta T/T \simeq 0.12$, exist in
nuclear collisions. It could mean therefore that we are dealing here
with some fluctuations existing in small parts of the system in
respect to the whole system (according to interpretation of \cite{L})
rather than with fluctuations of the event-by-event type in which,
for large multiplicity $N$, fluctuations $\Delta T/T = 0.06/
\sqrt{N}$ should be negligibly small \cite{FLUCT2}. This controversy
could be, in principle, settled by detailed analyses of the
event-by-event type. Already at present energies and nuclear targets
(and the more so at the new accelerators for heavy ions like RHIC at
Brookhaven, now commisioned, and LHC at CERN scheduled to be
operational in the year 2006) one should be able to check whether the
power-like $p_T$ distribution $dN(p_T)/dp_T$ occurs already at every
event or only after averaging over all events. In the former case we
would have a clear signal of thermal fluctuations of the type
mentioned above. In the latter case one would have for each event a
fixed $T$ value which would fluctuate from one event to another (most
probably because different initial conditions are encountered in a given
event).\\  

The proposed interpretation of $q$ leads immediately to the next
question: why and under what circumstances is it the gamma
distribution that describes fluctuations of the parameter $\lambda$?
To address it let us write the usual Langevin equation for the
stochastic variable $\lambda$ \cite{FP}:
\begin{equation}
\frac{d\lambda}{dt}\, +\, \left[\frac{1}{\tau}\, +\, \xi(t)\right]\,
\lambda\, =\, \phi\, =\, {\rm const}\, >\, 0 . \label{eq:LE}
\end{equation}
with damping constant $\tau$ and source term $\phi$. This term will
be different for the two cases considered, namely: 
\begin{equation}
\phi = \phi_{q<1}\, =\, \frac{1}{\tau}\left(\chi_0 -
                 \frac{\varepsilon}{\alpha'}\right) 
\qquad {\rm whereas}\qquad \phi = \phi_{q>1} = \frac{\chi_0}{\tau} .
             \label{eq:FIFI}
\end{equation}
For stochastic processes defined by the {\it white gaussian noise}
form of $\xi(t)$ \cite{FOOT3} one obtains the following Fokker-Plank
equation for the distribution function of the variable $\lambda$
\cite{FOOT4} 
\begin{equation}
\frac{df(\lambda)}{dt}\, =\, -\, \frac{\partial}{\partial \lambda}K_1\,
f(\lambda)\, +\, \frac{1}{2}\, \frac{\partial^2}{\partial \lambda^2}K_2\,
f(\lambda) , \label{eq:FPE}
\end{equation}
where the intensity coefficients $K_{1,2}$ are defined by
eq.(\ref{eq:LE}) and are equal to (cf., for example, \cite{ADT}): 
\begin{equation}
K_1(\lambda)\, =\, \phi\, -\, \frac{\lambda}{\tau}\, +\, D\, \lambda
\qquad {\rm and}\qquad 
K_2(\lambda)\, =\, 2\, D\, \lambda^2 . \label{eq:KK}
\end{equation}
From it we get the following  expression for the distribution
function of the variable $\lambda$:
\begin{equation}
f(\lambda)\, =\, \frac{c}{K_2(\lambda)}\, \exp\left[\, 2\,
\int^{\lambda}_0 d\lambda'\, \frac{K_1(\lambda')}{K_2(\lambda')}\, \right]
\label{eq:EF} 
\end{equation}
which is, indeed, a gamma distribution in variable $1/\lambda$,
\begin{equation}
f(\lambda)\, =\, \frac{1}{\Gamma(\alpha)}\, \mu\, 
 \left(\frac{\mu}{\lambda}\right)^{\alpha-1}\, \exp\left( -\,
\frac{\mu}{\lambda} \right) , \label{eq:FRES}
\end{equation}
with the constant $c$ defined by the normalization condition,
$\int^{\infty}_0 d(1/\lambda) f(1/\lambda) = 1$ and depending on two
parameters:  
\begin{equation}
\mu(\varepsilon)\, =\, \frac{\phi_q(\varepsilon)}{D} \qquad {\rm and}
\qquad \alpha_q\, =\, \frac{1}{\tau\, D} ,\label{eq:PAR}
\end{equation}
with $\phi_q = \phi_{q>1,q<1}$ and $\alpha_q = (\alpha, \alpha')$
for, respectively, $q>1$ and $q<1$. This means that we have
obtained eq. (\ref{eq:QG1}) with $\omega = \frac{1}{\tau D}$ and,
therefore, the parameter of nonextensivity $q$ is given by the parameter
$D$ and by the damping constant $\tau$ describing the {\it white
noise}.\\    

The above discussion rests on the stochastic equation (\ref{eq:LE}).
To comment on its possible origin let us turn once more to
fluctuations of temperature \cite{L,LS,FLUCT1,FLUCT2} discussed
before, i.e., to $\lambda = T$. Suppose that we have a thermodynamic
system, in a small (mentally separated) part of which the temperature
fluctuates with $\Delta T \sim T$. Let $\lambda(t)$ describe
stochastic changes of the temperature in time. If the mean temperature of
the system is $\langle T\rangle = T_0$ then, as result of
fluctuations in some small selected region, the actual temperature
equals $T' = T_0 - \tau \xi(t) T$. The inevitable exchange of heat
between this selected region and the rest of the system leads to the
equilibration of the temperature and this process is described by the
following equation \cite{LLH} 
\begin{equation}
\frac{\partial T}{\partial t}\, -\, \frac{1}{\tau}
            \, (T'\, -\, T)\, +\, \Omega_q = 0 \label{eq:HC}
\end{equation}
which is, indeed, of the type of eq. (\ref{eq:LE}) (here
$\Omega_{q<1} = \frac{\varepsilon}{\tau \alpha'}$ and $\Omega_{q>1} 
= 0$).\\

In this way we have recovered eq. (\ref{eq:LE}) and clearly
demonstrated the plausibility of our proposition. Notice the presence
of the internal heat source in the above equation in the $q<1$ case.
It has a sense of dissipative transfer of energy from the region where
(due to fluctuations) the temperature $T$ is higher. It could be any
kind of convection-type flow of energy; for example, it could be
connected with emission of particles from that region. The heat
release given by $\varepsilon/(\tau\alpha')$ depends on $\varepsilon$
(but it is only a part of $\varepsilon$ that is released). In the
case of such energy release (connected with emission of particles)
there is additional cooling of the whole system. If this process is
sufficiently fast, it could happen that there is no way to reach a
stationary distribution of temperature (because the transfer of 
heat from the outside can be not sufficient for the development of the
state of equilibrium). On the other hand (albeit this is not our case
here) for the reverse process we could face the "heat explosion"
situation (which could happen if the velocity of the exothermic
burning reaction grows sufficiently fast; in this case because of
nonexistence of stationary distribution we have fast nonstationary
heating of the substance and acceleration of the respective
reaction).\\    

It should be noticed that in the case of $q<1$ the temperature does
not reach stationary state because, cf. Eq. (\ref{eq:MV}),
$\langle 1/T \rangle\, =\, 1/(T_0 - \varepsilon/\alpha')$,
whereas for $q>1$ we had $<1/T> = 1/T_0$. As a consequence the
corresponding L\'evy distributions are defined only for $\varepsilon
\in(0, T_0\, \alpha'$) because for $\varepsilon \rightarrow
T_0\alpha'$, $<T>\rightarrow 0$. Such asymptotic (i.e., for
$t/\tau \rightarrow \infty$) cooling of the system ($T\rightarrow 0$)
can be also deduced form Eq. (\ref{eq:HC}) for $\varepsilon
\rightarrow T_0\alpha'$.\\

Our explanation, being tied to specific examples (especially to
the example of the temperature fluctuations) differs from other works
in which $L_{q\neq1}(\varepsilon)$ is shown to be connected with
$L_{q=1}(\varepsilon)$ by the so called Hilhorst integral formula
(the trace of which is our eq. (\ref{eq:GF})) \cite{FOOT2,C} but
without discussing the physical context of the problem. Our original
motivation was to understand the apparent success of Tsallis
statistics (i.e., the situations in which $q>1$ or, possibly also
$q<1$) in the realm of high energy collisions. It should be stressed
that in this way we have addressed the interpretation of only very
limited cases of applications of Tsallis statistics. They belong to
the category in which the power laws physically appear 
as a consequence of some continuous spectra within appropriate
integrals. It does not touch, however, a really {\it hard} case of
applicability of Tsallis statistics, namely when {\it zero} Lyapunov
exponents are involved \cite{FOOT5}. Nevertheless, this allows us to
interpret some nuclear collisions data in terms of fluctuations of
the inverse temperature, providing thus an important hint to the
origin of some systematics in the data, understanding of which is
crucial in the search for a new state of matter, the Quark Gluon
Plasma \cite{ALQ,FLUCT2}.\\ 

\section{Final remarks}

The are also other imprints of nonextensivity which we shall only
mention. One is connected with recent analysis \cite{JR} of the
equilibrium distribution of heavy quarks in Fokker-Planck dynamics.
It was demonstrated that thermalization of charmed quarks in a QGP
proceeding via collisions with light quarks and gluons results in a 
spectral shape which can be described only by the Tsallis distrubution
\cite{FOOTR}. 
On the other hand in \cite{II} the quantum scattering processes
(such as $\pi N$ and $\pi A$) scatterings were analysed using
Tsallis-like entropies and strong evidence for the nonextensivity
were found there when analysing the experimental data on the
respective phase shifts.
On the boundary of really high
energy collisions is the very recent application of the nonextensive
statistics to the nuclear multifragmentation processes \cite{GPPT}.
The other examples
 do not refer to Tsallis thermostatistics directly,
nevertheless it can be demonstrated that they are, at least
approximately, connected to it. We would like to refer here to a recent
attempt to study, by using the formalism of quantum groups, the so called
Bose-Einstein correlations between identical particles observed in
multiparticle reactions \cite{BECQ} and also works on intermittency and
multiparticle distributions using the so called L\'evy stable
distributions \cite{INTER}. They belong, in some sense, to the domain
of nonextensivity because, as was shown in \cite{TQ}, there is 
close correspondence between the deformation parameter of quantum
groups used in \cite{BECQ} and the nonextensivity parameter $q$ of
Tsallis statistics and there is also connection between Tsallis
statistics and L\'evy stable distributions \cite{ZEM}. Some traces of
the possible nonextensive evolution of cascade type hadronization
processes were also searched for in \cite{UWWC}. The quantum group
approach \cite{BECQ,TQ} could probably be a useful tool when studying 
delicate problem of interplay between QGP and hadrons produced from
it. It is plausible that description in terms of $q$-deformed bosons
(or the use of some kind of interpolating statistics) would lead to
more general results than the  simple use of nonextensive mean
occupation numbers $<n>_q$ discussed above (for which the only known
practical description is limited to small deviations from
nonextensivity only). \\ 

To the extend to which self-organized criticality (SOC) is connected
with nonextensivity \cite{T} one should also mention here a very
innovative (from the point of view of high energy collision)
application of the concept of SOC to such processes \cite{MENG}. \\

To summarize, it has been demonstrated that multiparticle processes
bear also some signs of nonextensivity observed in other branches of
physics, which shows up only as small deviations from the 
expected behaviour. These deviations were already explained by
invoking some additional mechanisms and, because of this, the use of
$q$-statistics is not so popular or known in this field as in others
discussed in \cite{T}. The advantage of the use of $q$-statistics is 
probably best seen from the information theoretical point of view.
The new parameter $q$ can be regarded then as a kind of
compactification of all processes responsible for the actual
nonextensivity into one single number \cite{BII}.
This is also the point of view expressed in \cite{QFT} where the new
approach to quantum field theory based on Lorentzian, instead of
Gaussian, path integrals has been proposed. It would allow to account
for the possible deviations of pure stochasticity in a similarly most
economical way when one introduces a single new parameter. This is,
however, so far unexplored domain of research.\\  

Acknowledgements: The partial support of Polish Committee for
Scientific Research (grant 621/E-78/SPUB/CERN/P-03/DZ4/99) is
acknowledged. One of us (G.W.) is very grateful to Prof. P.Grigolini and
Prof. C.Tsallis, organizers of the Complexity 2000 Workshop, for the 
financial support and warm hospitality extended to him during 
the conference. We thank Dr. A.Petridis for critical reading of the
manuscript.\\


\newpage
\noindent
{\bf Figure Captions}\\

\begin{itemize}

 \item[{\bf Fig. 1}] Depth distribution of the starting points,
                     $dN(T)/dT$, of cascades in 
                     Pamir lead chamber. Notice the non-exponential
                     behaviour of data points (for their origin cf.
                     \cite{WWCR}) which can be fitted by Tsallis
                     distribution (\ref{eq:Q13}) with $q=1.3$. (This 
                     figure is reproduced from Fig. 1 of \cite{CR}).

 \item[{\bf Fig. 2}] The results for $p_T$ distribution 
                     $dN(p_T)/dp_T$: notice that
                     $q=1.015$ results describes also the tail of
                     distribution not fitted by the conventional
                     exponent (i.e., $q=1$). This figure is
                     reproduced from Fig. 3 of \cite{UWW}.

 \item[{\bf Fig. 3}] The example (Fig. 1 of \cite{BCM}) of transverse
                     momentum distributions for $e^+e^-$ annihilation
                     processes for differnt energies (shown in Figure).
                     The inset shows region of small values of $p_T$.
                     The dotted line shows curve of constant $T$ and
                     $q=1$. Other curves were obtained by fitting
                     corresponding data with essentially fixed 
                     $T\simeq 110$ MeV and $q$ growing fast with energy 
                     to stabilize at $\sim 90$ GeV at value $q=1.2$
                     (cf. \cite{BCM} for details; this figure is
                     reproduced from Fig. 1 of \cite{BCM}).

 \item[{\bf Fig. 4}] The examples of the most probable rapidity
                     distributions obtained by extending analysis
                     of \cite{MAXENT} (eq. (\ref{eq:MAXENT})) to the 
                     nonexponential ($q\neq 1$) distributions given
                     by eq. (\ref{eq:MAXENTq}). The object (fireball,
                     string,...) of mass $M=100$ GeV decays into $N$ 
                     secondaries of (transverse) mass $m_T=0.4$ GeV 
                     each. Figs. $(a)$ and $(b)$ show results for 
                     $N$ leading to {\it Feynman scaling} ($\beta =0$)
                     or {\it Feynman $q$-scaling} ($\beta_{q=1.3}=0$).
                     Fig. $(c)$ shows example of such $N$ that all
                     $\beta > 0$.
                    
 \item[{\bf Fig. 5}] Example of the nonextensivity in fluctuations: 
                     Fig. 1 of \cite{UWW} showing $\Phi$ - measure 
                     of the kaon multiplicity fluctuations
                     (in the $\pi^- K^-$ system of particles)
                     as a function of temperature for three values of
                     the pion chemical potential. The kaon chemical
                     potential vanishes. The resonances are neglected. 
                     $(a)$ - results of \cite{MN1} (in linear scale);
                     $(b)$ - our results for $q=1.015$.

\end{itemize}

\newpage
\begin{figure}[h]
\setlength{\unitlength}{1cm}
\begin{picture}(25.,16.5)
\includegraphics{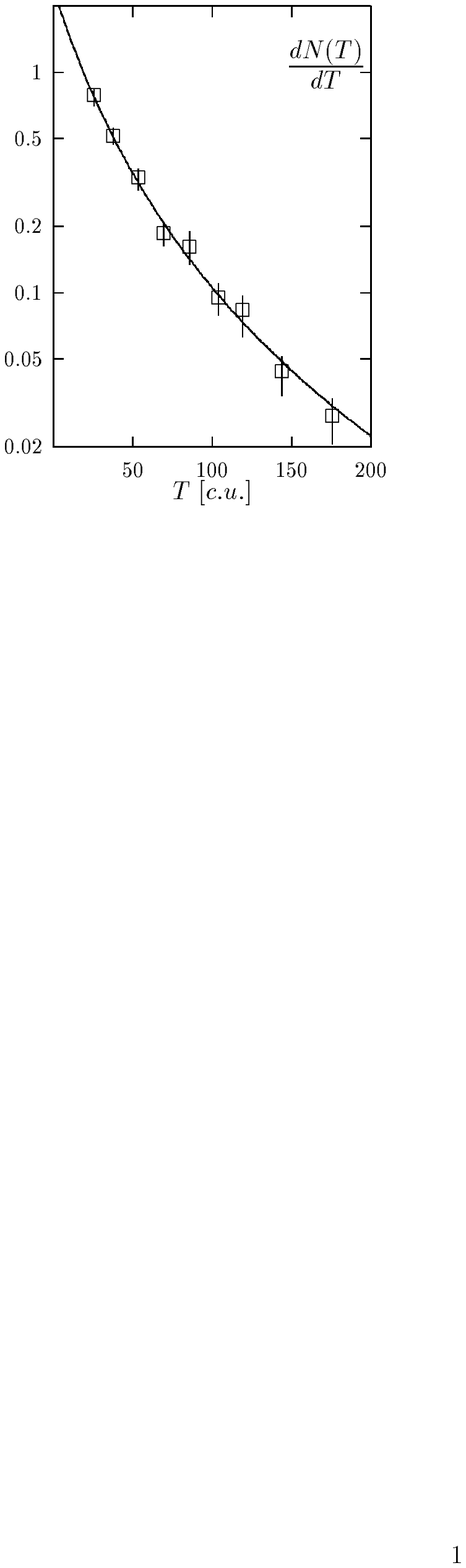}
\end{picture}
\end{figure}
\begin{center}
Fig. 1
\end{center}

\newpage
\begin{figure}[h]
\setlength{\unitlength}{1cm}
\begin{picture}(25.,16.5)
\includegraphics{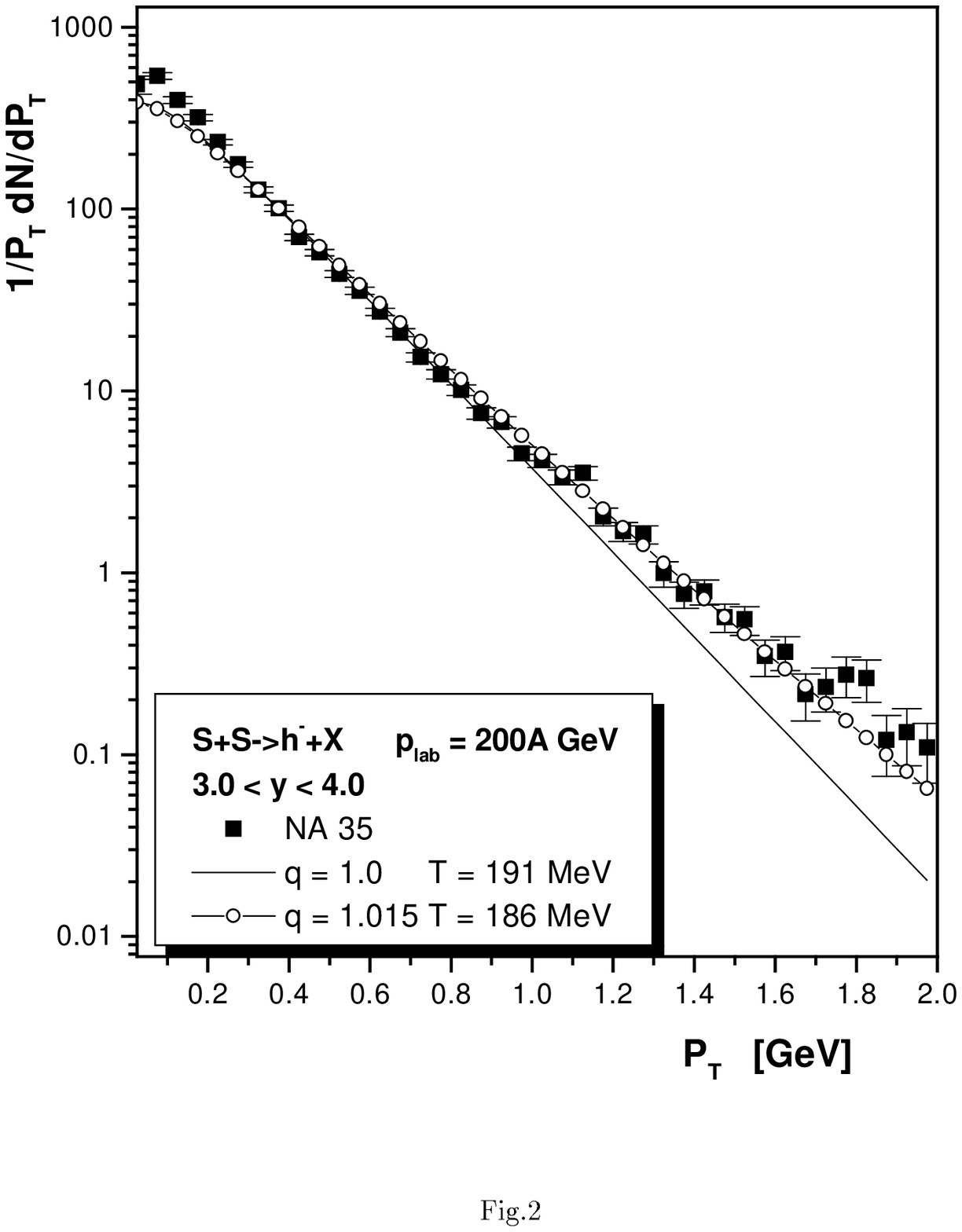}
\end{picture}
\end{figure}

\newpage
\begin{figure}[h]
\setlength{\unitlength}{1cm}
\begin{picture}(25.,16.5)
\includegraphics{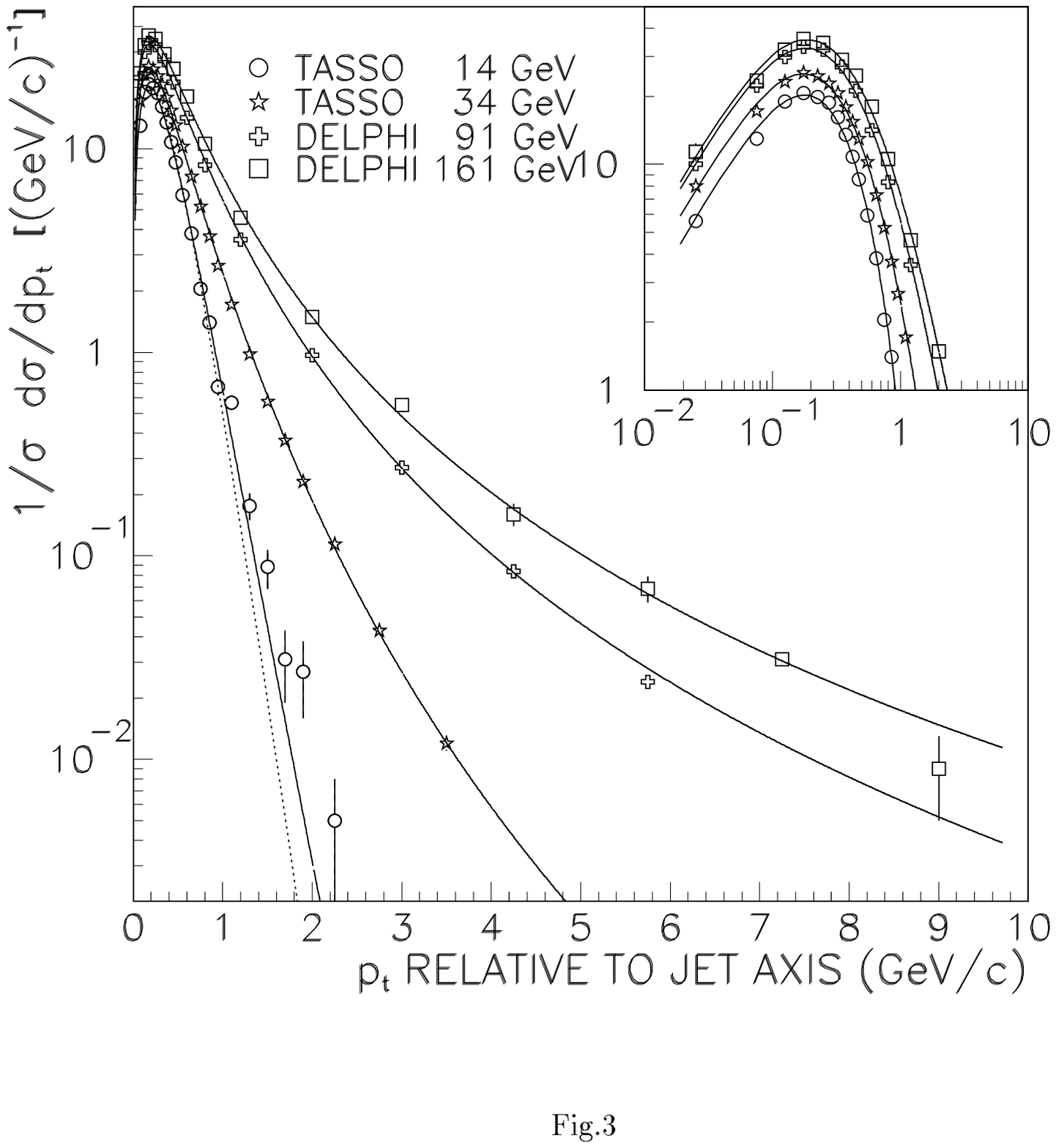}
\end{picture}
\end{figure}

\newpage
\begin{figure}[h]
\setlength{\unitlength}{1cm}
\begin{picture}(25.,16.5)
\includegraphics{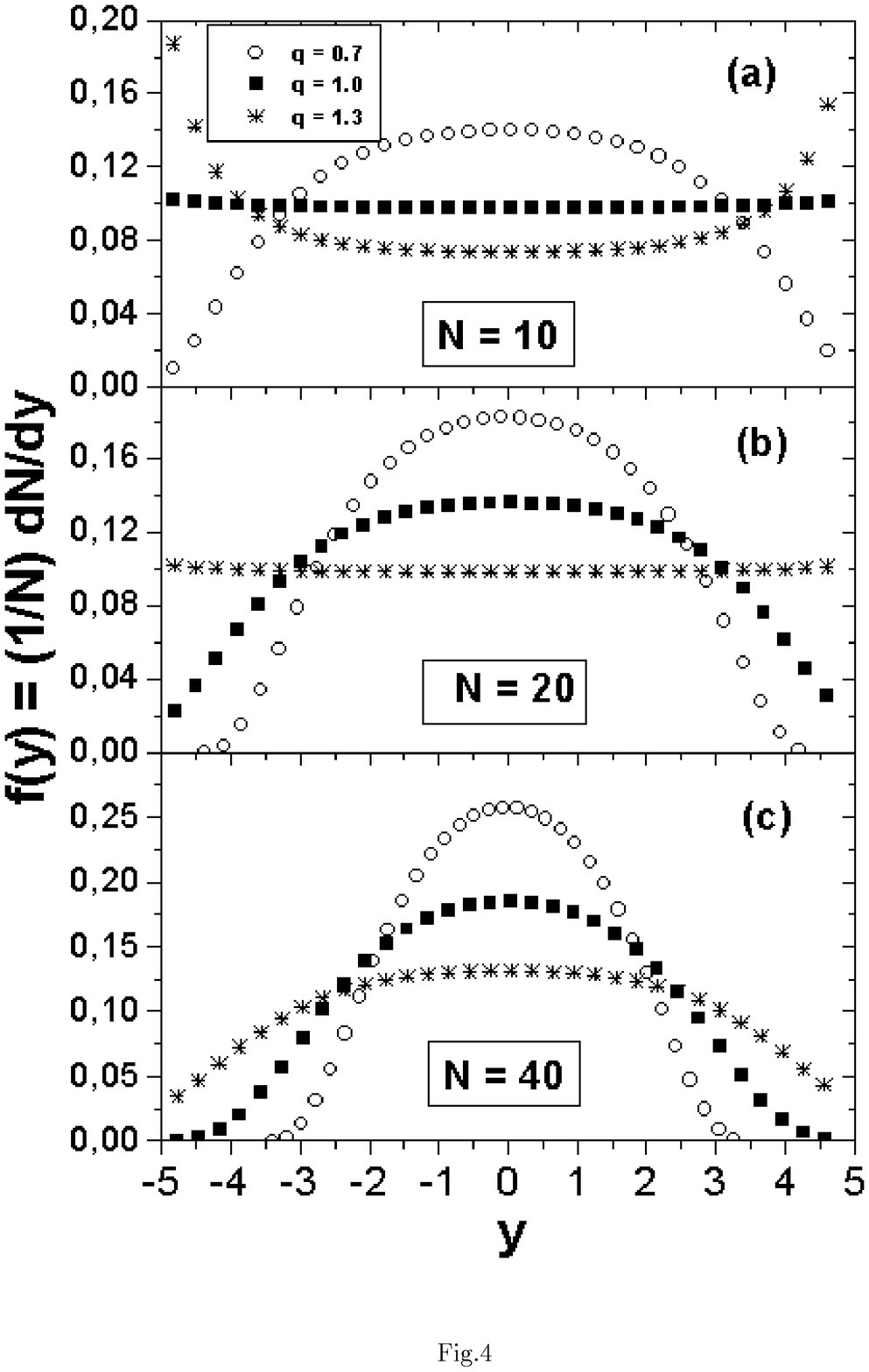}
\end{picture}
\end{figure}

\newpage
\begin{figure}[h]
\setlength{\unitlength}{1cm}
\begin{picture}(25.,16.5)
\includegraphics{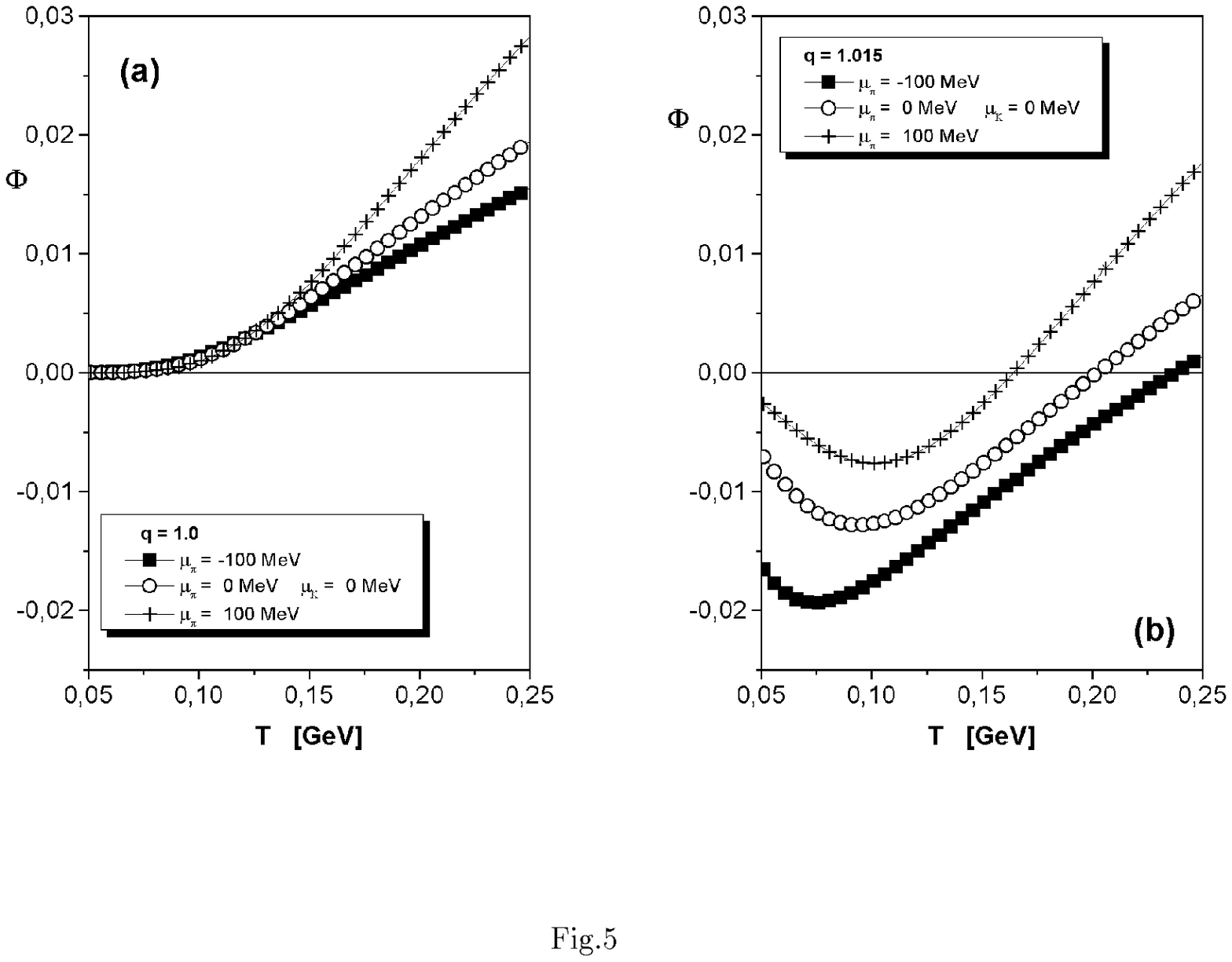}
\end{picture}
\end{figure}


\begin{thebibliography}{99}
 
 \bibitem{T} C.Tsallis, {\sl J. Stat. Phys.} {\bf 52} (1988) 479;
             for updated bibliography on this subject see 
             http://tsallis.cat.cbpf.br/biblio.htm. Its recent 
             summary is provided in the special issue of {\sl Braz. 
             J. Phys.} {\bf 29} (No 1) (1999) (available also at 
             http://sbf.if.usp.br/WWW\_pages/Journals/BJP/Vol29/Num1/index.htm). 
             
 \bibitem{FOOT0} This will be the only form used here, i.e., we 
                 shall not discuss (or use) the more general definition 
                 exploiting instead the normalized $q$-expectation
                 values in the form
                 $<A>_q = \int dx p(x)^q A(x)/\int dx p(x)^q$
                 (cf. \cite{T} and C.Tsallis, R.S.Mendes and A.R.Plastino,
                 {\sl Physica} {\bf A261} (1998) 534).

 \bibitem{CASC} Cf., for example, T.E.Harris, {\it The Theory of
                Branching Processes}, Springer, Berlin 1963;
                A. Ohsawa, {\sl Suppl. Prog. Theor. Phys.} {\bf 47} 
                (1971) 180.

 \bibitem{LANDAU} L.D. Landau and S.Z. Bilenkij, {\sl Nouvo Cim.
                  Suppl.} {\bf 3} (1956) 15 and {\it Collected Papers
                  of L.D. Landau}, ed. D. Ter Haar, Pergamon Press
                  (1965), p. 569. This model is still in use even
                  today for quick and straightforward estimations of
                  some quantities of interest \cite{MG}) 
 
 \bibitem{MG} M.Ga\'zdzicki, {\sl Z. Phys.} {\bf C66} (1995) 659.
 
 \bibitem{QGP} Cf., for example, proceedings of {\it Quark Matter'99},
               {\sl Nucl. Phys.} {\bf A525} (1999) and references
               therein.
 
 \bibitem{CR} G.Wilk and Z.W\l odarczyk, {\sl Nucl. Phys.} {\bf B} 
              ({\sl Proc. Suppl.}) {\bf A75} (1999) 191.

 \bibitem{BCM} I.Bediaga, E.M.F.Curado and J.M.de Miranda, {\sl Physica}
               {\bf A286} (2000) 156.
              
 \bibitem{ALQ} W.M.Alberico, A.Lavagno and P.Quarati, {\sl Eur. Phys.
               J.} {\bf C12} (2000) 499. 

 \bibitem{UWW} O.V.Utyuzh, G.Wilk and Z.W\l odarczyk, {\sl J. Phys.}
               {\bf G26} (2000) L39.

 \bibitem{WW} G.Wilk and Z.W\l odarczyk, {\sl Phys. Rev. Lett.}
              {\bf 84} (2000) 2770 and {\it Some remarks on the
              interpretation of degree of nonextensivity}, hep-ph/0002145.

 \bibitem{JR} D.B.Walton and J.Rafelski, {\sl Phys. Rev. Lett.} 
              {\bf 84} (2000) 31.

 \bibitem{II} M.L.D.Ion and D.B.Ion, {\sl Phys. Lett.} {\bf B482} (2000)
              57; {\sl Phys. Rev. Lett.} {\bf 83} (1999) 463;
              {\sl Phys. Rev.} {\bf E60} (1999) 5261 and references therein
              (cf. also contributed paper to this workshop).
              
 \bibitem{WWCR} G.Wilk and Z.W\l odarczyk, {\sl Phys. Rev.} {\bf D50}
                (1994) 2318. 

 \bibitem{FOOT1} One should keep in mind that there are also other
                 results from cosmic ray experiments, which presumably
                 could also be explained in this way, for example:
                 non-exponential character of absorption of the EAS
                 (Extensive Air Showers) in atmosphere or the so
                 called "Tien-Shan effect" which consists of the
                 existence of the long flying component in the
                 ionizing calorimeter with lead absorbers, cf.
                 references in \cite{WWCR}. 

 \bibitem{NONM} Cf., for example, S.Gavin, {\sl Nucl. Phys.} 
                {\bf B351} (1991) 561;
                H.Heiselberg and X.N.Wang, {\sl Phys/ Rev.} {\bf
                C53} (1996) 1892; 
                T.S.Bir\'o and C.Greiner, {\sl Phys. Rev. Lett.} 
                {\bf 79} (1997) 3138; 
                J.Rau, {\sl Phys. Rev.} {\bf D50} (1994) 6911; 
                S.Schmidt et al., {\sl Int, J. Mod. Phys.} {\bf E7} 
                (1998) 709 and {\sl Phys. Rev.} {\bf D59} (1999) 
                094005;
                Cs. Anderlik et al., {\sl Phys. Rev.} {\bf C59}
                (1999) 388 and 3309;
                V.K. Magas et al., {\sl Phys. Lett.} {\bf B459} 33
                (1999).
                Cf. also \cite{JR}. 
                J.Hormuzdiar, S.D.H.Hsu and G.Mahlon, {\it Particle
                Multiplicities and Thermalization in High Energy
                Collisions}, McGill/00-04 and OITS-686 preprints;
                arXiv:nucl-th/0001044. 
                 
 \bibitem{LUND} Cf., for example, B. Andersson, {\sl J. Phys.} 
                {\bf G17} (1991) 1507 or K.Werner, {\sl Phys. Rep.} 
                {\bf 232} (1993) 87 and references therein.

 \bibitem{MAXENT} G.Wilk and Z.W\l odarczyk, {\sl Phys. Rev.} {\bf D43}
                  (1991) 794.
                  
 \bibitem{UWWF} O.V.Utyuzh, G.Wilk and Z.W\l odarczyk, {\it Violation
                of the Feynman scaling law as a manifestation of
                nonextensivity}, presented at Chacaltaya Meeting on
                Cosmic Ray Physics, July 23-27, 2000, La Paz, Bolivia,
                to be published in {\sl Nuovo Cim.} {\bf C}; 
                hep-ph/0009165.
 
 \bibitem{UWWFa} Actually the problem of existence or not of Feynman
                 scaling and form of its possible violation is very
                 important in cosmic ray physics, cf. A.Ohsawa, {\sl
                 Prog. Theor. Phys.} {\bf 92} (1994) 1005; J.Wdowczyk
                 and A.W.Wolfendale, {\sl J. Phys.} {\bf G10} 
                 (1984) 257; {\bf G13} (1987) 411 and \cite{UWWF}.
 
 \bibitem{QSTAT} F.B\"uy\"ukkili\c{c}, D.Demirhan and A.G\"ule\~c, 
                 {\sl Phys. Lett.} {\bf A197} (1995) 209; 
                 U.Tirnakli, F.B\"uy\"ukkili\c{c}, D.Demirhan, 
                 {\sl Physica} {\bf A240} (1997) 657.

 \bibitem{GM} M.Ga\'zdzicki and S.Mr\'owczy\'nski, {\sl Z.Phys.} {\bf
              C54} (1992) 127. 

 \bibitem{MN} S.Mr\'owczy\'nski, {\sl Phys. Lett.} {\bf B439} (1998) 6.

 \bibitem{MNPHI} S.Mr\'owczy\'nski, {\it $\Phi$-measure of azimuthal
                 fluctuations}, nucl-th/9907099. For a generalization
                 of $\Phi$-measure see S.Mr\'owczy\'nski,
                 {\sl Phys. Lett.} {\bf B465} (1999) 8.

 \bibitem{MN1} S.Mr\'owczy\'nski, {\sl Phys. Lett.} {\bf B459} (1999) 13.

 \bibitem{FOOT6} One can argue that resonance production belongs in
                 our nonextensive philosophy already to the
                 nonextensive case being therefore responsible for
                 (at least a part of) the effect leading to  
                 a nonzero $|1-q|$. This is best seen inspecting
                 results of \cite{MN1} with resonances included,
                 which show that $\Phi$ in this case also changes
                 sign. The use of parameter $q$ is, however, more
                 general as it includes all other possible effects as
                 well. 

 \bibitem{L} L.D.Landau, I.M.Lifschitz, {\it Course of Theoretical 
             Physics: Statistical Physics}, Pergmon Press, New
             York 1958.

 \bibitem{LS} L.Stodolsky, {\sl Phys. Rev. Lett.} {\bf 75} (1995)
              1044.

 \bibitem{FLUCT1} T.C.P.Chui, D.R.Swanson, M.J.Adriaans,
                  J.A.Nissen and J.A.Lipa, {\sl Phys. Rev. Lett.} {\bf
                  69} (1992) 3005; C.Kittel, {\sl Physics Today} {\bf 5} 
                  (1988) 93; B.B.Mandelbrot, {\sl Physics Today} {\bf
                  1} (1989) 71; H.B.Prosper, {\sl Am. J. Phys.} {\bf 61}
                  (1993) 54; G.D.J.Phillires, {\sl Am. J. Phys.} {\bf
                  52} (1984) 629.                  
                  
 \bibitem{FLUCT2} E.V.Shuryak, {\sl Phys. Lett.} {\bf B423} (1998) 9 and 
                  S.Mr\'owczy\'nski, {\sl Phys. Lett.} {\bf B430} (1998) 9.

 \bibitem{FOOT2} This use of what is essentially 
                 the Mellin transformation has been discussed in
                 different contexts of Tsallis statistics in a
                 number of places, cf., for example, D.Prato, {\sl
                 Phys. Lett.} {\bf A203} (1995) 165; P.A.Alemany,
                 {\sl Phys. Lett.} {\bf A235} (1997) 452 or C.Tsallis
                 et al., {\sl Phys. Rev.} {\bf E56} (1997) R4922.  

 \bibitem{FP} N.G. van Kampen, {\it Stochastic Processes in Physics
              and Chemistry}, Elsevier Science Pub. B.V.,
              North-Holland, Amsterdam 1987 (Chapter VIII).

 \bibitem{FOOT3} It means that ensemble mean $\langle \xi(t)
                 \rangle\, =\, 0 $ and correlator (for sufficiently
                 fast changes) $\langle \xi(t)\, \xi(t + \Delta t)
                 \rangle\, =\, 2\, D\, \delta(\Delta t)$.
                 Constants $\tau$ and $D$ define, respectively, the
                 mean time for changes and their variance by means of
                 the following conditions: $\langle
                 \lambda(t)\rangle\, =\, \lambda_0\, \exp\left( -
                 \frac{t}{\tau} \right)$ and $\langle
                 \lambda^2(t=\infty)\rangle\, =\, \frac{1}{2}\, D\,
                 \tau$. Thermodynamical equilibrium is assumed here
                 (i.e., $t >> \tau$, in which case the influence of
                 the initial condition $\lambda_0$ vanishes and the
                 mean squared of $\lambda$ has value corresponding to
                 the state of equilibrium).

 \bibitem{FOOT4} Notice that our discussion (and results that follow) 
                 resembles (but only to some extent) approaches where 
                 Tsallis distributions were derived as exact solution 
                 of the standard Fokker-Planck equation, see for example, 
                 L.Borland, {\sl Phys. Lett.} {\bf A245} (1998) 67 and 
                 {\sl Phys. Rev.} {\bf E57} (1998) 6634. 
                 
 \bibitem{ADT} C.A.Ahmatov, Y.E.Diakov and A.Tchirkin, {\it Introduction
               to Statistical Radiophysics and Optics}, Nauka,
               Moscow, 1981 (in Russian).

 \bibitem{LLH} L.D.Landau and I.M.Lifschitz, {\it Course of Theoretical
               Physics: Hydrodynamics}, Pergamon Press, New York 1958 
               or {\it Course of Theoretical Physics: Mechanics of
               Continous Media}, Pergamon Press, Oxford 1981.

 \bibitem{C} C.f., for example, S.Curilef, {\sl Z.Phys.} {\bf B100} 
             (1996) 433 and references therein.

 \bibitem{FOOT5} See for example M.L.Lyra and C.Tsallis, {\sl Phys.
                 Rev. Lett.} {\bf 80} (1998) 53, C.Anteneodo and
                 C.Tsallis, {\sl Phys. Rev. Lett.} {\bf 80} (1998)
                 5313, and references therein. 

 \bibitem{FOOTR} We tend to regard this results as being yet another
                 example of the possible influence of fluctuations
                 discussed in the previous section. They could
                 originate, for example, from the finite (and 
                 fluctuating) space-time extension of the region
                 of QGP production. 

 \bibitem{GPPT} K.K.Gudima, A.S.Parvan, M.P.\l oszajczak and
                V.D.Toneev, {\it Nuclear Multifragmentation in the
                Non-extensive Statistics - Canonical Formulation},
                arXiv:nucl-th/0003025. 

 \bibitem{BECQ} D.V.Anchishkin, A.M.Gavrilik and N.Z.Iogorov, {\it
                Two-particle correlations from the $q$-boson viewpoint},
                CERN-TH/99-177 preprint, nucl-th/9906034; to be published
                in {\sl Eur. J. Phys.} {\bf C} (2000).                
 
 \bibitem{INTER} Cf., for example, R.Peschanski, {\sl Nucl. Phys.} 
                 {\bf B253} (1991) 225 or S.Hegyi, {\sl Phys. Lett.} 
                 {\bf B387} (1996) 642 (and references therein).

 \bibitem{TQ} C.Tsallis, {\sl Phys. Lett.} {\bf A195} (1994) 329.
              See also A.Lavagno and P.Narayana Swamy, {\sl Mod.
              Phys. Lett.} {\bf 13} (1999) 961 and {\sl Phys. Rev.}
              {\bf E61} (2000) 1218.

 \bibitem{ZEM} Cf. D.H.Zanette in review volume quoted in \cite{T},
               p. 108.

 \bibitem{UWWC} O.V.Utyuzh, G.Wilk and Z.W\l odarczyk, {\sl Czech J. 
                Phys.} {\bf 50/S2} (2000) 132.

 \bibitem{MENG} Meng Ta-chung, R.Rittel and Z.Yang, {\sl Phys. Rev.
                Lett.} {\bf 82} (1999) 2044 and C.Boros, Meng
                Ta-chung, R.Rittel, K.Tabelow and Z.Yang, {\it
                Formation of color-singlet gluon-clusters and
                inelastic diffractive scattering}, hep-ph/9905318, to
                be published in {\sl Phys. Rev.} {\bf D} (2000). 
                
 \bibitem{BII} Although in the examples discussed here the values of the
               parameter $q$ were quite close to unity, they can vary
               in quite large range in other applications, cf. for
               example, B.M.Boghosian, {\sl Phys. Rev.} {\bf E53}
               (1996) 4754; C.Anteodo and C.Tsallis, {\sl J. Mol. 
               Liq.} {\bf 71} (1997) 255 and \cite{II}.

 \bibitem{QFT} R.A.Treuman, {\sl Phys. Rev.} {\bf E57} (1998) 5150;
               {\sl Europh. Lett.} {\bf 48} (1999) 8 and {\sl Phys.
               Scripta} {\bf 59} (1999) 19, 204. See also R.S.Mendes
               in review volume quoted in \cite{T}, p. 66.

\end{thebibliography}
\end{document}